\documentclass[amsmath,amsfonts,twocolumn,superscriptaddress,prl]{revtex4-1}
\usepackage{graphicx}
\usepackage{epsfig}
\usepackage{bm}
\usepackage{dcolumn}
\usepackage{amsmath}
\usepackage{amssymb}
\usepackage{gensymb}
\usepackage{color}
\usepackage{textgreek}
\usepackage[varg]{txfonts}

\begin{document}

\title{Interaction Effects and Viscous Magneto-Transport in a Strongly Correlated 2D Hole System}

\author{Arvind Shankar Kumar} 
\affiliation{Department of Physics, Case Western Reserve University, 2076 Adelbert Road, Cleveland, Ohio 44106, USA} 

\author{Chieh-Wen Liu}, 
\affiliation{Department of Physics, Case Western Reserve University, 2076 Adelbert Road, Cleveland, Ohio 44106, USA} 

\author{Shuhao Liu}
\affiliation{Department of Physics, Case Western Reserve University, 2076 Adelbert Road, Cleveland, Ohio 44106, USA} 

\author{Loren N. Pfeiffer}  
\affiliation{Department of Electrical Engineering, Princeton University, Princeton, New Jersey 08544, USA} 
 
\author{Kenneth W. West}
\affiliation{Department of Electrical Engineering, Princeton University, Princeton, New Jersey 08544, USA}  
 
\author{Alex Levchenko}
\affiliation{Department of Physics, University of Wisconsin--Madison, Madison, Wisconsin 53706, USA} 
 
\author{Xuan P. A. Gao}
\affiliation{Department of Physics, Case Western Reserve University, 2076 Adelbert Road, Cleveland, Ohio 44106, USA} 

\begin{abstract}
Fermi liquid theory has been a foundation in understanding the electronic properties of materials. For weakly interacting two-dimensional (2D) electron or hole systems, electron-electron interactions are known to introduce quantum corrections to the Drude conductivity in the FL theory, giving rise to temperature dependent conductivity and magneto-resistance. Here we study the magneto-transport in a strongly interacting 2D hole system over a broad range of temperatures ($T$ = 0.09 to $>$1K) and densities $p=1.98-0.99\times10^{10}$ /cm\textsuperscript{2} where the ratio between Coulomb energy and Fermi energy $r_s$ = 20 -- 30. We show that while the system exhibits a negative parabolic magneto-resistance at low temperatures ($\lesssim$ 0.4K) characteristic of an interacting FL, the FL interaction corrections represent an insignificant fraction of the total conductivity. Surprisingly, a positive magneto-resistance emerges at high temperatures and grows with increasing temperature even in the regime $T \sim E_F$, close to the Fermi temperature. This unusual positive magneto-resistance at high temperatures is attributed to the collective viscous transport of 2D hole fluid in the hydrodynamic regime where holes scatter frequently with each other. These findings highlight the collective transport in a strongly interacting 2D system in the $r_s\gg 1$ regime and the hydrodynamic transport induced magneto-resistance opens up possibilities to new routes of magneto-resistance at high temperatures.  
\end{abstract}

\date{May 14, 2021}

\maketitle

\textbf{Introduction}. Much of recent interest in condensed matter physics has been focused on strongly correlated electronic systems that go beyond the famed Fermi Liquid (FL) theory. In FL paradigm, low temperature excitations of a many-electron system are characterized as independent quasiparticles with an interaction-renormalized effective mass and scattering properties, in a Fermi sea \cite{nozieres1997theory}. Through this quasiparticle picture, low-temperature properties of a majority of electronic systems where non-trivial band-structure effects do not play a role have been successfully understood. Non-FL behavior has been observed and studied in a variety of systems like  $d$- and $f$-electron compounds \cite{Stewart2001} and high-$T_c$ superconductors \cite{Hill2001}. Most recently, magic angle twisted bilayer graphene, a system of two graphene sheets aligned at a 1.1 degree relative angle, has shown similar evidence of non-FL behavior and a variety of interesting correlation-induced phases \cite{Cao2018,Cao2018a,Sharpe2019,Zondiner2020}.

One notable success of FL theory has been in describing and predicting transport properties of electronic systems, via considering the quantum properties of quasiparticles moving through the disorder potential landscape in a crystalline lattice where the effects of quantum interference or Coulomb interactions are accounted for perturbatively as corrections to the classical Drude conductivity. Hence for a disordered FL with sufficiently weak interactions, the longitudinal conductivity is found to be described by the well-known Drude model with quantum corrections arising from single-particle quantum interference effects like weak localization (WL), and electron-electron interactions (EEI) \cite{Altshuler1983,ALTSHULER1985}. The single particle quantum interference effect in 2D electronic systems is suppressed by a perpendicular magnetic field through time-reversal symmetry breaking and completely disappears in the strong-field regime $\omega_c\tau > 1$  ( $\omega_c=eB/m^*$ is the cyclotron frequency and $\tau$ is the momentum-relaxation time, $e$ is the electron charge and $m^*$ is the effective mass). The EEI corrections to conductivity are weakly affected by magnetic field only via Maki-Thompson term, which is present even in a nominally non-superconducting systems, however it is suppressed logarithmically in temperature and decays rapidly with increasing magnetic field similar to WL. As a result, it can be shown that the interplay of these factors leads to a negative parabolic magneto-resistance (MR) from the EEI \cite{Houghton1982,ALTSHULER1985,Gornyi2003,AL2009}. Negative parabolic MR has subsequently been observed in many 2D electronic systems such as GaAs/AlGaAs heterostructures \cite{Li2003,Bockhorn2011} and more recently in 2D materials like InSe \cite{Kumar2020} and epitaxial graphene \cite{Jobst2012}, further lending to the success and ubiquity of FL theory. However, since FL theory treats the Coulomb interactions between electrons in the many-particle system as a perturbation, it is strictly valid only for $r_s < 1$. Yet, FL theory is known to be empirically valid for many electronic systems beyond the small $r_s$ limit (e.g. at $r_s>1$ in 2D electrons \cite{Li2003}). There were also attempts to apply FL interaction correction theory to experimental systems into the regime where $r_s \gg $1 \cite{Proskuryakov2002,Noh2003} although this perturbative approach of FL theory may not be sufficient in the strongly interacting regime where anomalous transport phenomena such as the metal-insulator transition occur \cite{Abrahams2001, GaoRMP2010}. 

\begin{figure*}[t!]
\begin{center}
\includegraphics[width=0.95\textwidth]{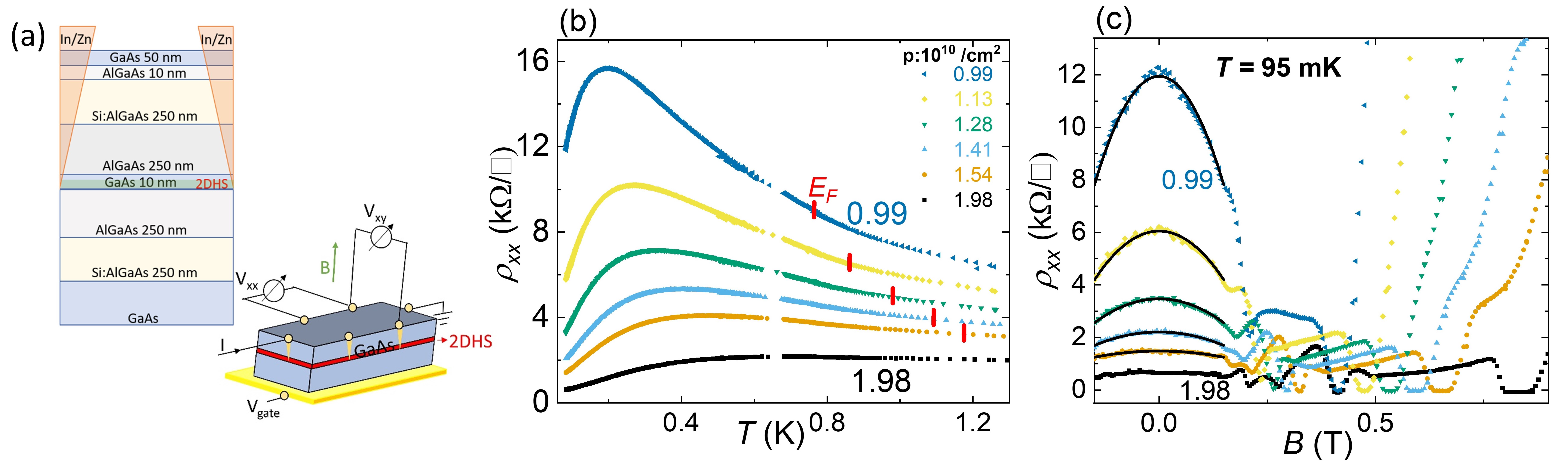}
\end{center}
\caption{\label{fig:epsart} (a) GaAs/AlGaAs 10 nm QW device structure (side view) and device schematic. (b) $\rho_{xx} (T)$ versus temperature $T$ for the 2DHS confined in QW, plotted for various hole densities $p$ listed in figure, in the range $0.99 -1.98 \times 10^{10}$ /cm\textsuperscript{2}. Values of $E_F$ (in units of $k_B$) for each density are marked in red on the curves. (c) $\rho_{xx} (T)$ versus magnetic field $B$, applied perpendicular to the 2DHS plane, at $T$=95 mK for the densities listed in (b). Black solid lines correspond to fits to $B^2$dependence.}
\label{MRdata}
\end{figure*}

More recently, FL analysis of electron/hole transport has been extended to the hydrodynamic regime, where quasiparticles interact with each other more frequently than with phonons, impurities or the sample boundaries \cite{Andreev2011,Crossno2016,Lucas2018,Levchenko2017,Mandal2020}. In this case, conductivity can no longer be described by the Drude model with perturbative interaction corrections, since it is an interaction dominated regime. In this case, the frequent electron-electron/hole-hole collisions have been shown to create behavior that is in analogy to a classical viscous fluid. Signatures of viscous behavior in electron and hole transport has been a popular direction of current research in different Dirac electron systems in which the disorder-limited carrier mean free path is very long such that the hydrodynamic regime can be accessed \cite{Polini2020,Lucas2018,Berdyugin2019,Crossno2016,Sulpizio2019,Moll2016,Gooth2018}.  It is worth noting that the Dirac systems exhibiting hydrodynamic behavior do not reside in the high $r_s$ regime and the breakdown of the standard FL quasiparticle transport is a result of the frequent scattering of carriers at elevated temperatures but not the dominance of Coulomb interaction energy in the system.

High mobility 2D carrier systems with low densities in semiconductor heterointerfaces is an exciting platform to discover/understand strong interactions induced breakdown of FL quasiparticle transport. Indeed, most of the aforementioned examples of materials exhibiting non-FL behavior are also systems having strong correlations. For 2D electron systems (2DES) or 2D hole systems (2DHS), the interaction effects are expected to be strong in the low density regime since in 2D, $r_s=1/(a^* \sqrt{\pi p})$ with $p$ as the electron or hole density and $a^*=\hbar^2 \epsilon/m^* e^2$ as the effective Bohr radius. Hence low density 2D systems with a low carrier concentration and/or large effect mass offer an interesting platform to test the limits of FL theory. Thanks to the tunable carrier density, and thus interaction strength, low density 2DES/2DHS with strong correlations have attracted great interest in studying phase transitions and novel quantum solid phases \cite{kravchenko1994,Yoon1999,Qiu2012,Knighton2014,Huang2018,Brussarski2018,Shashkin2019,Jang2017}. However, whether the conventional Boltzmann transport and quantum correction theories of FL can apply to 2DES/2DHS with $r_s\gg1$ has remained a subject of intense discussion \cite{GaoRMP2010}.

\begin{figure*}[t!]
\includegraphics[width=0.95\textwidth]{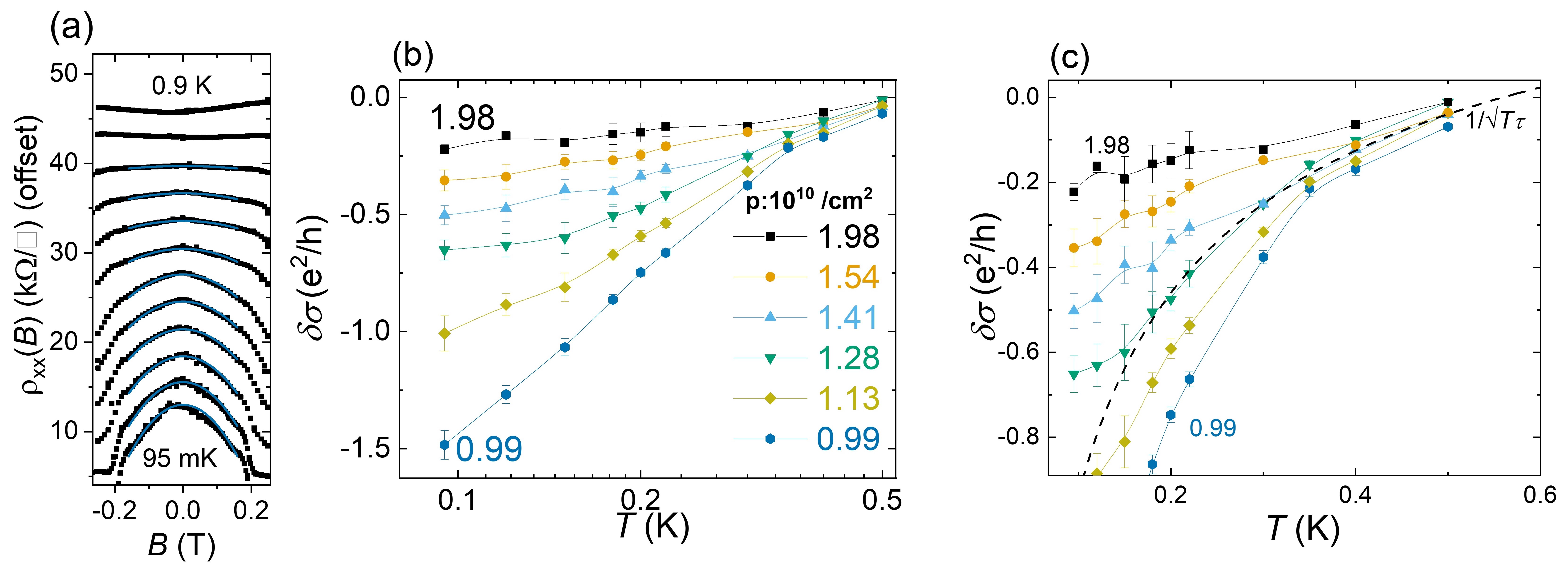}
\caption{\label{fig:epsart} (a)  $\rho_{xx}$ vs. $B$ at $p = 0.99\times10^{10}$ /cm\textsuperscript{2}, plotted for different temperatures from 95 mK to 0.9 K. Curves have been offset along the y-axis in the order of increasing temperature for clarity (from bottom – 95 mK, 0.12, 0.15, 0.18, 0.22, 0.30, 0.35, 0.40, 0.50, 0.70, 0.90K), and plotted for -0.25T $< B <$ 0.25T to clearly shown negative parabolic MR. Blue lines correspond to fits to $B^2$ dependence. (b) Correction to conductivity $\delta\sigma (T)$ extracted from MR data using Eq.(1), plotted versus temperature for the densities listed in figure. (c) $\delta\sigma (T)$ data (as shown in (b) ) zoomed in to show the high temperature range. Dashed line corresponds to fit to $1/\sqrt{T\tau}$ dependence for the $p=1.41$ over the $T\geq 0.3$ K range. }
\label{EEIcorrection}
\end{figure*}

\textbf{Negative Parabolic MR at Low Temperatures}. Our experiments were performed on modulation doped p-type GaAs  Quantum Wells (QWs) with 10 nm thickness. The samples were grown on (100) GaAs wafer (Samples 1-3) or (311) GaAs wafer (Sample 4) with Al\textsubscript{0.1}Ga\textsubscript{0.9}As barriers and symmetrically placed Si delta doping layer 200 nm away from the QW, creating a high mobility, ultra-dilute 2DHS in the GaAs QW (see Figure \ref{MRdata}(a) for device structure). The hole density of the samples without gating was $p \sim 2$ (for this paper, hole densities are in units of $10^{10}$ /cm\textsuperscript{2}) and the low-temperature mobility was $\mu \sim 1-2\times10^5$ cm\textsuperscript{2}/Vs over the density range explored. The carrier density is tuned ($p\sim 1-2$) using a back-gate placed approximately 150 \textmu m away from the 2DHS, so that screening of Coulomb interactions due to the back gate is negligible (see Methods for further fabrication and measurement details).

In Figure \ref{MRdata}(b), we plot  $\rho_{xx} (T)$ for our GaAs 2DHS sample (for clarity, all data presented in main text are from Sample 1 except where explicitly stated), where we observe a non-monotonic behavior which crosses over from insulating-like behavior at high temperature to metallic behavior with decreasing temperature. This sign change in $ \partial\rho/\partial T$ has been previously observed over this density range in samples of similar mobility \cite{Yoon1999,Abrahams2001,Goble2014} and is a subject of active research in the context of 2D metal-insulator transition (in other samples, we have observed the turnover to insulating behavior throughout the whole $T$-range at the lowest densities - plotted in Figure S1, Supplementary Info). Figure \ref{MRdata}(c) shows longitudinal magneto-resistivity $\rho_{xx} (B)$ - the focus of our study - at 95 mK where, in addition to dips corresponding to quantum Hall states at higher perpendicular magnetic field $B$, we observe a clear negative parabolic MR at $B \lesssim 0.2$ T over the entire density range. The temperature evolution of the observed parabolic MR in Figure \ref{MRdata}(c) is plotted for $p=0.99$ in Figure \ref{EEIcorrection}(a). Similar data for other densities and samples are plotted in Figures S2 and S3 in Supplementary Information.

While the Drude model of non-interacting electrons predicts zero MR,  a $B$-independent correction $\delta\sigma$ to the longitudinal Drude conductivity combined with the Drude magneto-conductivity, which has a $1/(1+(\mu B)^2)$ dependence, leads to a parabolic MR - whose magnitude is determined by $\delta\sigma$ as given by \cite{Altshuler1983,Houghton1982,Gornyi2003,AL2009}  
\begin{equation}
\rho_{xx}=\frac{1}{\sigma_0 }\left[1-(1-(\omega_c \tau)^2)\frac{\delta\sigma}{\sigma_0} \right]. 
\label{MReq}
\end{equation}
As mentioned earlier, such corrections are often attributed to perturbative EEI/HHI effects, which is indeed the dominant factor in weakly interacting FLs. However, for our interaction-dominated system, $r_s \gg 1$, it is not clear that EEI/HHI can be treated as perturbative corrections. In spite of this, we can still extract a correction $\delta\sigma$ to the conductivity  according to Equation \ref{MReq} from the observed parabolic MR data and compare to relevant theories. 

The extracted $\delta\sigma$ from the observed negative parabolic MR similar to Figure \ref{EEIcorrection}(a) are plotted versus temperature $T$ for various densities in Figure \ref{EEIcorrection}(b) and (c). It is worth noting that during the fitting process, $\delta\sigma$ is the only parameter fitted from the curvature of the $B^2$ dependent $\rho_{xx}(B)$ at a given $T$ after we substitute $\omega_c \tau$ with $\mu B$ and $\sigma_0$ with $p e \mu$ which cancels the unknown $\mu$ or $\tau$. We observe that $\delta\sigma (T)$ gives a negative correction that tends to make the system more insulating, and the magnitude of the correction increases as we lower the density. Also, the $T$-dependence of the correction $\delta\sigma (T)$ seems to follow the $\ln T$ dependence at lower temperatures (Figure \ref{EEIcorrection}(b)) and diminishes towards zero as $T$ increases (Figure \ref{EEIcorrection}(c)), qualitatively similar to the expectations of quantum corrections due to HHIs in the FL theory \cite{Altshuler1983,Zala2001,Gornyi2003}. However, the negative sign (and insulating-like nature) of $\delta\sigma (T)$ is opposite to the strong metallic trend of the system within this temperature range. Moreover, a detailed analysis of $\delta\sigma (T)$ using FL theories covering both diffusive and ballistic regimes (see Methods) yields unphysical behavior of the FL parameter $F_0^\sigma$ despite the fact that the $T$-dependence may be described by the $\ln T$ and $1/\sqrt{T}$ behavior as the FL interaction correction theories expect in the diffusive and ballistic regimes respectively \cite{Zala2001,Gornyi2003}. We conclude that perturbative quantum corrections to the quasiparticle transport in the FL picture is not a dominant factor in the overall conductivity of the experimental system at large $r_s$.

 \begin{figure}[t]
 \begin{center}
 \includegraphics[width=0.5\textwidth]{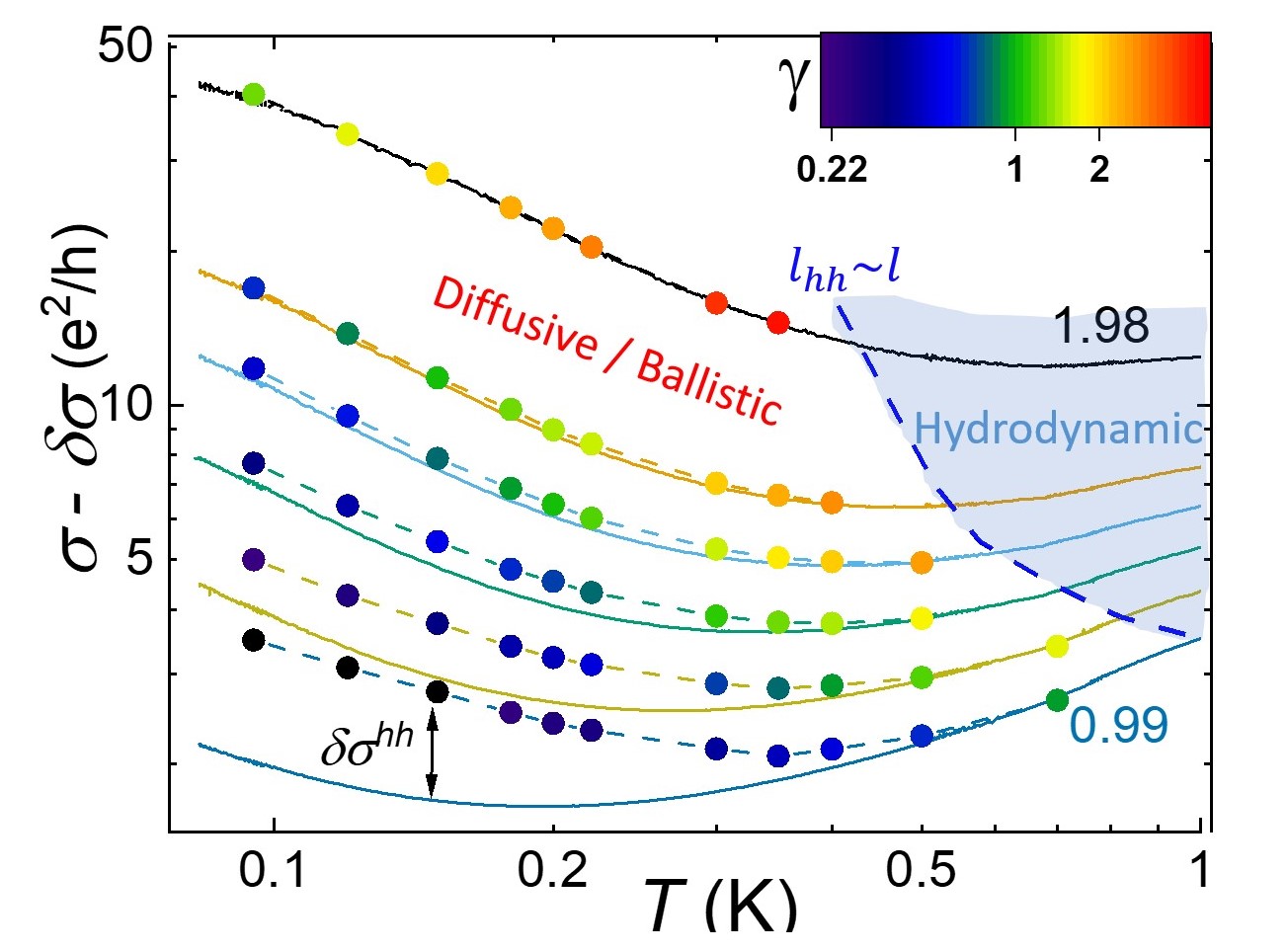}
\end{center}
\caption{\label{fig:epsart} $\sigma - \delta\sigma$, conductivity without HHI correction,  plotted versus $T$ for the densities listed in Figure \ref{MRdata} and \ref{EEIcorrection}. The colour of each point indicates $\gamma=\frac{k_{B}T}{\hbar/\tau}$ values, as shown in the inset colour map, and colour of the dashed line along the points indicates corresponding density values as in Figure \ref{EEIcorrection}. The solid lines are a plot of measured $\sigma$ vs $T$ for the corresponding densities (same data as in Figure \ref{MRdata}(b)). The blue dashed line indicates the temperature above which $l_{hh} \gtrsim l$ for the respective curves and the shaded region indicates the corresponding hydrodynamic transport regime where Drude conductivity with corrections no longer applies (see Methods section). }
\label{Rawcon}
\end{figure}

\textbf{Magneto-transport and Hydrodynamic Behavior at High Temperatures}. In light of this, we plot $\sigma-\delta\sigma$ versus $T$, in Figure \ref{Rawcon} to examine the behavior of the system's raw conductivity without the quantum correction. The raw conductivity shows a strong metallic temperature dependence at low temperatures ($T\lesssim 0.4$K), which cannot be attributed to quantum corrections as discussed above. Classical single-particle scattering effects like hole-phonon scattering is also not relevant in this temperature and density range (see Figure S6 in Supplementary Information for scattering length calculations). In fact, at the lowest density $p=0.99$, we see that $\delta\sigma$ constitutes a relatively large fraction ($\sim 30-40 \%$) of the conductivity at the lowest temperature, but this fraction consistently reduces at higher density, and is close to zero at $p=1.98$. At the same time, at high temperatures ($\sim$0.4 - 1K) where the quantum correction effects are essentially zero, the system exhibits an unusual insulating like temperature dependence (also see the data showing decreasing resistivity at high $T$ in Figure \ref{MRdata}(b)). In this regime, since the system's carrier density is low and temperature $T\sim E_F$ -the Fermi energy (in units of $k_B$) - it was speculated that the increasing $\sigma$ or decreasing $\rho$ with $T$ is due to the $T$-dependent viscosity of a correlated fluid in the 'semi-quantum' regime, in analogy to Helium-3 fluid at $T\sim E_F$ \cite{spivak2005}. Yet there has been no further experimental test of this idea of understanding the transport in high $r_s$ 2D system using hydrodynamics which we address below. 

\begin{figure*}[t]
\begin{center}
\includegraphics[width=0.95\textwidth]{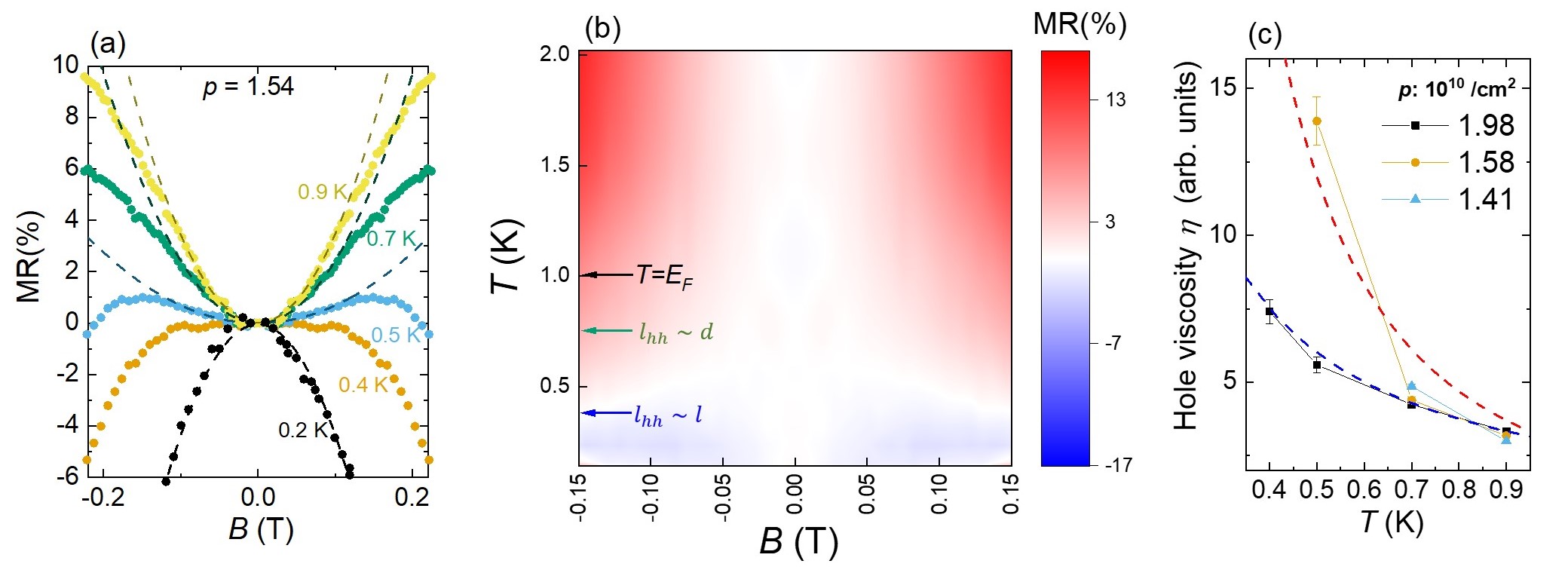}
\end{center}
\caption{\label{fig:epsart} (a) MR ($=\Delta\rho(B)/\rho(0) $ (\%) )  versus $B$ at $p=$1.54 plotted for $T=$ 0.2-0.9 K. Dashed lines are fits to $B^2$ dependence of low-field data (-0.1 T $< B <$ 0.1T) for corresponding temperatures. (b) Color plot of MR(\%) versus $B$ and $T$ (for Sample 4 at $p=1.13$) upto $T=2$ K. Blue and green arrows indicate temperatures above which $l_{hh} \lesssim l$ and $l_{hh} \lesssim d$ respectively, and the black arrow marks $T=E_F$ (See Figure S8 in Supplementary Information for additional details). (c) Temperature dependence of hole viscosity – extracted from fitting the MR at high $T$ as in (a) using Eq.\ref{Hydeq}, at different hole densities listed in figure. The solid lines are a guide to the eye. Blue and red dashed lines correspond to $\eta \propto \frac{1}{T}$ and $\eta \propto \frac{1}{T^2}$ dependencies, respectively.
}
\label{Hydro}
\end{figure*}

In electronic materials residing in the hydrodynamic transport regime, quasiparticles collide with each other and loose coherence before scattering with impurities or phonons. This is typically seen at relatively high temperatures, where the quasiparticle-quasiparticle scattering length (hole-hole scattering in our case) $l_{hh}$ is shorter than the mean free path $l$. This transport regime is termed the hydrodynamic regime because the electrons are seen to behave like a viscous fluid thanks to $l_{hh}$ being the shortest length scale and the electron system's momentum being conserved. Recent studies have found evidence for interesting correlated hydrodynamic transport effects in graphene \cite{Lucas2018,Berdyugin2019,Crossno2016}, PdCoO\textsubscript{2} \cite{Moll2016} and Weyl semimetal WP\textsubscript{2} \cite{Gooth2018}. A few recent theoretical works have also predicted specific magneto-resistance effects as a signature for the hydrodynamic transport which has not been explored experimentally \cite{Andreev2011,Levchenko2017,Mandal2020}. 
Interestingly, for our strongly correlated 2DHS system with high mobility, we find that the criterion $l_{hh}< l$ for the hydrodynamic regime is satisfied at higher temperatures ($T>0.4$K) where the $\sigma$ increases with $T$, as marked in Figure \ref{Rawcon} (further discussion of $l_{hh}$ and $l$ in Supplementary Info, Section 5 - Figure S6). We now turn our attention to the MR in this high temperature hydrodynamic regime where $l_{hh} \lesssim l$. Surprisingly, in the high-$T$ regime, we find that the MR does not disappear as in the standard classical Drude theory which predicts zero MR. Instead, a positive MR emerges at low $B$ and grows stronger as $T$ increases and this positive MR also appears to follow a parabolic $B$-dependence as shown in Figure \ref{Hydro}(a) for $p=1.54$.  It is remarkable that the positive MR in the hydrodynamic regime keeps increasing with temperature even at temperatures higher than the Fermi temperature, as shown in Figure \ref{Hydro}(b) for $p=1.13$ in a different sample.  (Additional data showing the switching from negative MR to positive parabolic MR in the hydrodynamic regime at high $T$ can be seen in Figure S3, and S7, S8 in the Supplementary Info).  

A MR that grows stronger with increasing temperature is unusual, especially in the regime $T\sim E_F$ where the classical Drude behavior (i.e. zero MR) is expected (the estimated Fermi energy is marked on Fig.1b for different densities). To understand this, we consider a contribution to conductivity and MR from the viscous fluid-like transport of holes in the hydrodynamic regime. In recent theoretical model \cite{Levchenko2017}, it was predicted that such viscosity-effects in a 2DES/2DHS with long-range disorder lead to a positive parabolic MR, whose temperature dependence is due to the temperature dependence of the viscosity of the electron/hole fluid. This regime occurs above a characteristic temperature determined from the condition $l_{hh}(T)\lesssim d$, where $d$ is the distance to the dopant spacer layer, that marks the onset of hydrodynamic behavior. The predicted MR of the the form 
\begin{equation}
\rho(T)=\rho_0 (T) +\frac{ \ln(L/d) }{2\hbar c^2 p \eta(T)} B^2
\label{Hydeq}
\end{equation}
where $L\gg \{l,1/k_F\}$ is a large spacial length scale, $\rho_0 (T)$ is the field-independent component of resistivity and $\eta (T)$ is the hole viscosity. The temperature evolution of the hole viscosity can hence be extracted from the positive parabolic MR in the high-$T$ hydrodynamic regime, seen in Figure \ref{Hydro}(a) for $p=1.54$, using Equation \ref{Hydeq} (Data from other densities and sample are plotted in Figure S7 and S8 in Supplementary Information). In Figure \ref{Hydro}(b), we see that the onset of this positive hydrodynamic MR coincides with the temperature range where $l_{hh} \sim l$ - marked by the blue arrow ($l_{hh} \sim d$ is also marked in the figure). Additionally, we also see from this figure that this positive MR continues to grow in the regime where $T>E_F$, contrary to classical Drude MR (see Figure S8 in Supplementary Information for further details).  In Figure \ref{Hydro}(c), we plot the extracted hole viscosity at a few different densities for the main sample discussed in the paper (the viscosity of sample 4 shown in Figure \ref{Hydro}(b) are presented in Fig.S9 in the supplementary Information). 

Figure \ref{Hydro}(c) shows a decreasing trend in the temperature dependence of hole viscosity (a similar trend was also observed over a larger temperature range in an additional sample - see Figure S8 and S9 in Supplementary Info). This $T$-dependence of viscosity extracted from the hydrodynamic MR alone without using other information follows the insulating-like decreasing $\rho(T)$ over this high-temperature range, as seen in Figure \ref{MRdata}(b) and Figure S1 and appears to be consistent with the viscosity explanation \cite{spivak2005} for this puzzling insulating-like temperature dependent $\rho(T)$ of high-mobility 2D carriers with large $r_s$ in the semi-quantum regime where temperature is high. Within the hydrodynamic description of transport one expects viscosity $\eta(T)\propto 1/T^2$ and $\eta(T)\propto 1/T$ in the $T<E_F$ and $T>E_F$ regimes, respectively (see Method section). Our extracted viscosity follows $\propto 1/T^b$ with $1<b<2$ is consistent with the fact the temperature range where the hydrodynamic MR emerges spans both $T<E_F$ and the semi-quantum regime $T\sim E_F$. The physical mechanism for the decreasing temperature dependence of resistivity ($\partial\rho/\partial T<0$) in a viscous electron/hole fluid is known as the Gurzhi effect, and is a prominent signature of hydrodynamic behavior in conducting systems \cite{dejong1995,Polini2020}. It has been observed in electrons in graphene, where a  temperature dependence of electron viscosity similar to in Figure \ref{Hydro}(c) has been shown to agree with predictions of many-body theory \cite{KrishnaKumar2017}. In our low density 2D system, it is known that $\rho(T)\propto 1/T$ in the semi-quantum regime $T\sim E_F$ \cite{Goble2014}, again consistent with the $\rho(T)$ being dominated by the viscosity effect with $\eta(T)\propto 1/T$ in this regime (see Method section).

\textbf{Discussion and conclusion}. The data in Figure \ref{EEIcorrection} is a signature of magnetic field independent quantum corrections to conductivity at low temperatures, as anticipated in standard FL theory of quasiparticle transport. However, the behavior of extracted conductivity correction contradicts the overall metallic conductivity behavior over this temperature range (Figure \ref{Rawcon}) and analysis in the interaction correction theories of FL  \cite{Gornyi2003}, does not result in a consistent $F_0^\sigma$ prediction. At higher temperatures, we also observe a positive MR that increases with $T$ (seen clearly in Figure \ref{Hydro}(b) up to the $T>E_F$ regime), contrary to standard FL interaction corrections and conventional MR that is usually suppressed at higher $T$.

In Figure \ref{Hydro} - along with the corresponding discussion - we see a resolution of this novel positive MR, along with the insulating-like $\rho (T)$ at higher temperature, through the consideration of viscous MR in the hydrodynamic regime and analysis of this effect, according to Ref. \cite{Levchenko2017}. Therefore, we believe that we have reached a consistent picture for both the unusual decreasing $\rho(T)$ and positive MR in the high $T$ regime through the hydrodynamic viscosity effect. It remains to be seen if further theories extending the hydrodynamics description to lower temperatures where the standard FL quantum interaction correction theories fail can quantitatively explain the strong metallic behavior of the system at low temperatures. One early idea was that the metallic conduction at low $T$ is due to the viscosity of the FL in the presence of Wigner crystal bubbles in the quantum degenerate regime \cite{spivak2005}.

In conclusion, we have observed the MR in a 2D hole system in GaAs/AlGaAs in a strongly correlated ($r_s=20-30$) limit switching from negative to positive parabolic behavior when the temperature is raised above the hydrodynamic regime where holes scatter with each other before impurities or phonons. At low temperatures where the MR is negative, we have found that the extracted correction to conductivity $\delta\sigma$ shows an opposite trend to the observed metallic conductivity over this temperature range, and constitutes  an insignificant fraction of the total conductivity. Thus the system's metallic conductivity at low temperatures cannot be accounted for by interaction corrections in conventional FL theory or single-particle scattering effects. In the high temperature hydrodynamic regime, we discover a positive MR due to viscous transport of 2D holes which reconciles with the decreasing resistivity in zero field as predicted in the Gurzhi effect. The unusual hydrodynamic MR that becomes stronger at higher temperature opens the door to a new way of understanding magneto-transport phenomena and inducing magneto-resistances at high temperatures in strongly correlated electronic systems.

\subsection*{Methods}

\subsubsection*{Material growth and device fabrication}

Our experiments were done on modulation doped p-type GaAs 10 nm Quantum Wells (QWs), which were grown by Molecular Beam Epitaxy on (100) or (311) GaAs wafer with Al\textsubscript{0.1}Ga\textsubscript{0.9}As barriers and symmetrically placed Si delta doping layer 200 nm away from the QW (see Figure \ref{MRdata}(a) for device structure). 
To make contacts to the heterostructure, Indium solder was prepared with $\sim$5\% Zn added to promote diffusion through the heterostructure layers, and solder spots were placed on the surface to define a 5mm $\times$ 3mm Hall bar. The samples were annealed at 425 \degree C for 4 min in a Rapid Thermal Annealer to diffuse the metal through to the 2DHS and 50 \textmu m diameter gold wires were soldered onto the spots. The sample was placed on a Au bottom gate (50 nm Au deposited on a Si substrate) using cryogenic vacuum grease. The distance between the gate and the 2DHS is approximately 150 \textmu m, so screening of Coulomb interactions due to the back gate is negligible.

\subsubsection*{Low Temperature magnetotransport measurements}

The sample was loaded onto a Dilution Refrigerator insert of a Quantum Design PPMS He-4 cryostat for low-temperature transport measurements. Magnetic fields upto 1.1T were applied using a 9T superconducting magnet for MR measurements. Measurements of $\rho_{xx} (T)$ from 95 mK to 900 mK were performed using the standard lock-in technique over the hole density range  $p=0.99-1.98\times 10^{10}/cm^2$ ($r_s=20-40$) by applying gate voltages from 0 to 70V and a measurement current of 5 nA to prevent hole heating. Densities at each gate voltage were extracted from the positions of observed Quantum Hall dips in $\rho_{xx}(B)$ (see, e.g., Figure  \ref{MRdata}(c)). MR measurements were done at different temperatures over the same density range by scanning the magnetic field from -1.1T to 1.1T or vice versa (no shift in MR behavior was observed between opposite sweep directions).

\subsubsection*{Fermi Liquid Interaction Correction Analysis of $\delta\sigma$}

Having extracted the corrections to Drude conductivity from MR (see Figure \ref{EEIcorrection} ), we examine the extracted $\delta\sigma (T)$ behavior with reference to interaction corrections in FL theory. Even though our 2DHS is probed in an interaction dominated regime whereas the following theory considers interactions as perturbative corrections, it is still worth making this comparison since most previous work on this topic use interaction corrections to analyze data close to this regime \cite{Proskuryakov2002,Noh2003}. Through this comparison, we will be able to establish whether the 2DHS in this regime can be described as a FL with interaction corrections, or a new description is required for this strongly interacting system. We note that Landau's paradigm of describing an interacting electron system using interaction parameters $F^\sigma$ and $F^\rho$ is valid up to arbitrarily strong $r_s$. 

For a FL with sufficiently weak EEI/HHIs, the value of $\gamma=\frac{E_{MC}}{E_{MR}}=\frac{k_{B}T}{\hbar/\tau}$, where mean free time $\tau$ is 
related to the Drude conductivity as $\sigma_0=pe^2\tau/m^*$, distinguishes different regimes of transport in our 2DHS. The $\gamma\ll1$ corresponds to the low $T$ regime of quasiparticles diffusing through a disordered lattice with quantum corrections arising from single particle interference or Coulomb interactions, while $\gamma \gtrsim 1$ corresponds to the high temperature ballistic regime where quasiparticle-quasiparticle scattering effects become relevant over the time scale of $\tau$. The $\gamma$ values for our system varies from 0.2-5 over the density and temperature range, with $\gamma\gtrsim1$ for $T\gtrsim0.4$K - $\gamma$ values for different densities and temperatures are indicated by the color of the points in Figure \ref{Rawcon}. For all our subsequent calculations, $\tau$ is calculated from the resistivity $\rho$ at the lowest temperature in our study (i.e. 95 mK).

To interpret our data in the interaction correction picture, we compare the temperature dependence of $\delta\sigma$  (which has been determined from MR data independent of $\sigma (T)$ - Figure \ref{EEIcorrection}(b)) to Gornyi \textit{et al.} \cite{Gornyi2003}, who have derived FL theory predictions for the temperature dependence of EEI/HHI correction to Drude conductivity for different types of disorder potentials and arbitrary $\gamma$ values, building on work by Zala \textit{et al.} \cite{Zala2001} who showed that interaction corrections to conductivity in both the ballistic and diffusive regimes have the same underlying physical origin. For the long-range smooth disorder relevant to our modulation doped samples, the HHI correction is given by 
\begin{equation}
\delta\sigma^{hh} (T)=\frac{e^2}{2\pi^2 \hbar} (G_F(\gamma)+G_H(\gamma))  
\label{GMeq}
\end{equation}
Here, $G_F$ is the Fock (exchange interaction) term, and $G_H$ is the Hartree term \cite{Gornyi2003}. For $\gamma\ll1$ (diffusive regime), $G_F(\gamma \ll 1)=-\ln(\gamma)+constant$ while $G_H$ undergoes strong FL renormalization to obtain $G_H(\gamma \ll 1)=(3(1-[\ln(1+F_0^\sigma)/F_0^\sigma]))\ln(\gamma)$. Here, $F_0^\sigma$ is the interaction parameter introduced in FL theory, which quantifies the strength of the FL renormalized quasiparticle interactions. For $\gamma\gtrsim 1$ (ballistic regime), $G_F(\gamma \gtrsim 1)=-(c_0/2)(\gamma)^{-1/2}$ and $G_H$ again undergoes FL renormalization to obtain $G_H(\gamma \gtrsim 1)=-(c_0/2)[3F_0^\sigma/(1+F_0^\sigma)](\gamma)^{-1/2}$, with $c_0 \simeq 0.276$. 

The logarithmic temperature dependence of $\delta\sigma$ observed in Figure  \ref{EEIcorrection}(b) at low temperatures $T<0.4 $K, clearly seen for $p=0.99,1.13$, is indicative of the diffusive regime as expected by $\gamma$ values ($<$1) over this temperature range. A logarithmic increase is also observed in the Hall coefficient as temperature is lowered (Figure S4 in Supplementary Information), consistent with predictions of FL interaction correction's temperature dependence in the diffusive regime \cite{Houghton1982}. However, we find that predictions of $F_0^\sigma$ extracted from MR and Hall effect are not consistent with each other (see Figure S5 in Supplementary Information).

In Figure \ref{EEIcorrection}(c), we take a closer look at $\delta\sigma (T)$ at higher temperatures $T \gtrsim $ 0.4K. In this high temperature regime, we observe the emergence of a positive MR which is seen to be parabolic at small magnetic fields (-0.1T $\lesssim B \lesssim$ 0.1T), which reverts to negative parabolic MR at higher fields for $T\simeq 0.4$K but remains positive at $T \gtrsim 0.7$K. We emphasize that this positive MR tends to become stronger with increasing temperature which is unusual and cannot be accounted for in the interaction correction theory given in Ref. \cite{Gornyi2003}. Therefore this positive MR appears to be an effect distinct from the negative parabolic MR. Its discussion is pursued in other sections which examine the hydrodynamic origin of this behaviour (see Figure \ref{Hydro} and corresponding discussion, and Figure S6 for scattering length calculations). To extract $\delta\sigma$ from the negative parabolic MR at higher temperatures where the low field positive MR coexists, we exclude the low magnetic field range where these two effects appear to compete (see Figure S7(a) in Supplementary Information for a clear depiction of these two competing effects) . 

Thus extending our FL interaction correction analysis of $\delta\sigma (T)$ to the ballistic regime, we find that $\delta\sigma (T)$ data may indeed be consistent with the $1/\sqrt{T\tau}$ dependence predicted for $T\gtrsim 0.4$K over the whole density range (Figure \ref{EEIcorrection}(c)), if we exclude the growing positive MR at higher temperatures. However, even in this case, we find that fitting the data to the ballistic limit of Equation \ref{GMeq} results in an unphysical $F_0^\sigma<-1$ over the entire density range. These facts, along with  previous discussions of $\delta\sigma (T)$ indicate that FL theory of HHI interaction corrections is insufficient to describe the system's transport despite the fact that there are some FL characteristics.

\subsubsection*{Fermi Liquid Hydrodynamic Theory of $\rho(T,B)$}

In the collision-dominated regime resistivity of the strongly correlated electron or hole liquid can be described in hydrodynamic terms. Thus it can be expressed in terms of pristine dissipative coefficients of the fluid, as well as some thermodynamics quantities. Following Refs. \cite{spivak2005,Andreev2011,Levchenko2017} and adopting the doping-layer model with the spatially uncorrelated dopants, which is relevant to our samples, the resistivity is found in the form 
\begin{align}\label{rho-hydro}
&\rho_0(T)=\rho_{\text{Th}}+\rho_{\text{Visc}}+\rho_{\text{MR}}, \nonumber\\
& \rho_{\text{Th}}=\frac{1}{16\pi e^2}\frac{T}{\kappa pd^2}\left(\frac{\partial s}{\partial\ln p}\right)^2,\quad \rho_{\text{Visc}}=\frac{3}{32\pi e^2}\frac{(\eta+\zeta)}{p^3d^4}.    
\end{align}
Here the first term $(\rho_{\text{Th}})$ is governed by thermal fluxes in the fluid and scales inversely proportionally to thermal conductivity $\kappa$. 
It is also proportional to the square of the entropy density variation. The second term $(\rho_{\text{Visc}})$ is due to viscous stresses in the fluid that scales with the sum of first-$\eta$ (shear) and second-$\zeta$ (bulk) viscosities. The last term represents MR per definition given earlier in Eq. \eqref{Hydeq}. The result expressed by Eq. \eqref{rho-hydro} is applicable to both weakly correlated ($r_s\sim1$) and strongly interacting ($r_s\gg1$) regimes. In both cases of hydrodynamic formalism, all microscopic details of interaction are captured by the temperature dependence of $\kappa(T)$, $\eta(T)$, $\zeta(T)$, and $s(T)$. These coefficients are well known in the FL theory of weakly interacting system at $r_s\sim1$, however, there are no microscopic results applicable to strongly interacting regime when $r_s\gg1$. In order to bypass this difficulty in extracting predictive statements from Eq. \eqref{rho-hydro} concerning temperature and main parametric interaction dependence of resistivity we rely on experimental facts. The key quantity of interest is the quantum lifetime $\tau_{hh}$ of the 2DHS that determines collision mean free path. Tunneling spectroscopy measurements by Eisenstein \textit{et al.} \cite{Eisenstein2007} reveal its temperature dependence to be consistent with the Fermi liquid theory predictions, $\tau^{-1}_{hh}\propto T^2$, for samples at least up to $r_s = 11$. To establish the pre-factor at large $r_s\gg1$, one should recognize that the energy associated with short scale rearrangements of particle positions is nonperturbatively renormalized. It involves quantum tunneling and is on the order of the spin-exchange constant $J$. We note that an instanton calculus in the Wigner crystal regime suggests $J\propto\exp(-\alpha\sqrt{r_s})$, where $\alpha$ is a number of order unity \cite{Roger1984}. Based on this arguments and also dimensional reasoning of the kinetic theory we conjecture $\kappa(T)\sim J^2/T$ and $\eta(T)/(m^*p)\sim J^3/pT^2$ for $T<E_F$. The final ingredient we need concerns entropy of the liquid $s(T)$. First, we note that at temperatures below the Debye energy, $T\ll\Omega_D\sim \sqrt{r_s}E_F$, plasmons give negligible contribution to the entropy. Thus the density of thermodynamic excitations is dominated by electron spins and two level systems. Both types of excitations involve tunneling, but there is significant difference between the two contributions. Indeed, in contrast to two level systems, spin exchange processes do not lead to rearrangements of electron positions. Because of the electrostatic energy cost associated with spatial rearrangement of electrons the density of two level systems is estimated to be $\sim Tp^{3/2}/e^2$ at $T\ll\Omega_D$. 
Therefore in a parametrically wide window of temperatures thermodynamic properties of the hole liquid are expected to be dominated by the spin exchange processes. In this regime the entropy per particle dependence may be written as a universal function of the ratio of the temperature to the spin exchange constant $s=s(T/J)$. In this picture one can simply relate $\partial s/\partial\ln p$ in Eq. \eqref{rho-hydro} to the heat capacity of the fluid $c_v=T(\partial s/\partial T)$, namely $\partial s/\partial\ln p=-c_v(\partial\ln J/\partial \ln p)$. By combining all this reasoning together we arrive at the following estimates from Eq. \eqref{rho-hydro}
\begin{align}
&\rho_{\text{Th}}/\rho_Q\sim (r_s/pd^2)(T/J)^4,\nonumber \\ 
&\rho_{\text{Visc}}/\rho_Q\sim(1/(pd^2)^2)(J^3/T^2E_F),\nonumber \\ 
&\rho_{\text{MR}}/\rho_Q\sim(T/J)^2(\omega^2_c/JE_F),
\end{align}
where $\rho_Q$ is the quantum of resistance and the extra factor of $r_s$ in $\rho_{\text{Th}}$ arises because $\ln J\propto\sqrt{r_s}$. The hydrodynamic approach is applicable as long as the mean free path $l_{hh}\sim v\tau_{hh}$ is shorter than $d$. This implies that the onset of hydrodynamic behavior requires $T>J/\sqrt{k_Fd}$. For our samples we estimate $k_Fd\sim 5$, a conservative estimate for the spin exchange gives $J\simeq \mathrm{Ry}^{*}r^{-5/4}_{s}e^{-\alpha\sqrt{r_s}}$ in the units of effective Rydberg energy, $\mathrm{Ry}^*=(m^*/m\epsilon^2)\mathrm{Ry}$, with $\alpha\sim 1$,  
$\epsilon$ being the dielectric constant of the host material, and the range of experiments is mostly within the domain $J<T<E_F$, where above assumptions are assumed to be valid.  

These expressions enable us to make several concrete statements and connect them to experimental observations. (i) The interplay of thermal and viscous contributions leads to the non-monotonic temperature dependence of the zero-field resistivity $\rho_0=\rho_{\text{Th}}+\rho_{\text{Visc}}$. The thermal and viscous term $\rho_{\text{Th}}$ and $\rho_{\text{Visc}}$ is an increasing/decreasing function of $T$ respectively and may describe the non-monotonic $\rho(T)$ observed in Figure \ref{MRdata}(b) and the literature, see Ref. \cite{GaoRMP2010}. However, at very strong interactions, $r_s\gg1$, more considerations are required. For instance, the picture of microemulsion phases suggests island formation of Wigner crystal phase located on top of the self-consistent potential \cite{spivak2005}. These macroscopic impurities lead to an additional viscous resistivity that can be estimated from the Stokes formula, $\rho_{\text{St}}\sim (N/e^2p^2)\eta$, where $N$ is the number of islands per unit area and we suppressed the weak logarithmic dependence of the resistance of a single island on its size. The ratio of the Stokes term to the viscous contribution in the fluid region is temperature independent and is given by $\rho_{\text{St}}/\rho_{\text{Visc}}\sim Npd^4$. For small concentration of the islands the bulk term is expected to dominate over Stokes resistivity mechanism. (ii) Within the onset of hydrodynamic regime, viscous term should dominate and lead to negative differential thermal resistance $\partial\rho/\partial T<0$, which is the hallmark of Gurzhi hydrodynamic effect.  Eq. \eqref{rho-hydro} remains valid in the semi-quantum regime, $E_F<T<\Omega_D$, albeit viscosity is expected to be inversely proportional to the temperature, $\eta\propto 1/T$, whereas the thermal conductivity is linear in the temperature $\kappa\propto T$. Thus the thermal diffusivity should become temperature independent and the $\rho_{\text{Th}}$ term should diminish with further increase of temperature. (iii) MR is positive and expected to follow quadratic temperature dependence, which gives a direct measure of fluid viscosity. Our data does not enable us to make a definite statement about $\eta(T)$ since our temperature range covers both $T<E_F$ and $T\sim E_F$. But Figure \ref{Hydro}(c) unambiguously indicates a power-law $\eta(T)\propto 1/T^b$ with $2\geq b\geq1$. This feature is consistent with the FL's viscosity $\eta\propto 1/T^2$ and $1/T$ in these two regimes which are observed in our samples, see Figure \ref{MRdata}(b). Therefore, combining all the arguments above, the temperature and magnetic field dependent transport of 2DHS with large $r_s$ here in the high-$T$ regime can be well understood in terms of hydrodynamic transport.  

\textbf{Acknowledgements}. We thank Anton Andreev, Steven Kivelson, and Boris Spivak for fruitful discussions and prior collaboration on related topics that shaped our understanding of systems at large $r_s$. XPAG acknowledges the financial support from NSF (DMR-1607631). The work of AL at UW-Madison was supported by the National Science Foundation CAREER Grant No. DMR-1653661.

\textbf{Competing Interests}. The authors declare no competing interests.

\textbf{Additional Information}. Supplementary Information is available for this paper.
Correspondence and requests for materials should be addressed to X.P.A.G. at xuan.gao@case.edu

\bibliography{references}

\end{document}